\documentclass{article}

     \PassOptionsToPackage{numbers, compress}{natbib}

\usepackage[final]{neurips_2024_ml4ps}

\usepackage[utf8]{inputenc} 
\usepackage[T1]{fontenc}    
\usepackage{hyperref}       
\usepackage{url}            
\usepackage{booktabs}       
\usepackage{amsfonts}       
\usepackage{nicefrac}       
\usepackage{microtype}      
\usepackage{xcolor}         
\usepackage{graphicx}
\usepackage{amsmath}
\usepackage{printlen}
\usepackage{wrapfig}
\usepackage{lipsum}
\usepackage{subcaption}
\usepackage{caption}

\usepackage{wrapfig}

\title{Deep Multimodal Representation Learning for Stellar Spectra}

\author{%
  Tobias Buck, Christian Schwarz\\
  Interdisciplinary Center for Scientific Computing, Heidelberg University, \\
  Im Neuenheimer Feld 205, D-69120 Heidelberg \\
  \texttt{tobias.buck@iwr.uni-heidelberg.de}, \texttt{b.schwarz@stud.uni-heidelberg.de}
}

\begin{document}

\maketitle

\begin{abstract}
Recently, contrastive learning (CL), a technique most prominently used in natural language and computer vision, has been used to train informative representation spaces for galaxy spectra and images in a self-supervised manner. Following this idea, we implement CL for stars in the Milky Way, for which recent astronomical surveys have produced a huge amount of heterogeneous data. Specifically, we investigate Gaia XP coefficients and RVS spectra. Thus, the methods presented in this work lay the foundation for aggregating the knowledge implicitly contained in the multimodal data to enable downstream tasks like cross-modal generation or fused stellar parameter estimation. We find that CL results in a highly structured representation space that exhibits explicit physical meaning. Using this representation space to perform cross-modal generation and stellar label regression results in excellent performance with high-quality generated samples as well as accurate and precise label predictions.
\end{abstract}

\section{Motivation and Related Work}
Over the past few decades, large-scale astronomical surveys of Milky Way stars have proliferated, including missions like Gaia \citep{Brown2021}, APOGEE \citep{2016AN....337..863M}, GALAH \citep{Buder2019,Buder2021}, and LAMOST \citep{lamost}, with upcoming efforts like 4MOST \citep{4most}. Each survey typically develops its own pipeline, requiring post-hoc calibration to reconcile discrepancies across different data sets \citep[e.g.][]{recio-blancoGaiaDataRelease2023}. This highlights the need for techniques capable of integrating data from diverse observational setups to maximize scientific output from heterogeneous datasets.
Recently, machine learning (ML) has enabled cross-survey analysis by harmonizing diverse datasets. For instance, AspGap \cite{liAspGapAugmentedStellar2023} aligns Gaia XP data with APOGEE, enabling the use of the APOGEE pipeline on Gaia observations. Similarly, ASTROCLIP \cite{lanusseAstroCLIPCrossModalPreTraining2023} generates low-dimensional vector representations of multi-band images and spectra from the Dark Energy Spectroscopic Instrument (DESI), effectively mapping objects to shared latent spaces for cross-modal analysis.
Moreover, \cite{leungAstronomicalFoundationModel2024} leverage a Transformer-based model trained on Gaia XP spectra and APOGEE stellar parameters to facilitate information transfer across different observation sets, exemplifying the emerging trend of foundation models in astronomy. Recent works in astrophysics also reflect the growing adoption of multimodal ML methods \cite{zhangMultimodalCelestialObject2023,hongPhotoRedshiftMMLMultimodalMachine2023,gaoDeepMultimodalNetworks2023,liuMFPIMDeepLearning2023,weiIdentificationBlueHorizontal2023,ait-ouahmedMultimodalityImprovedCNN2023,mishra-sharmaPAPERCLIPAssociatingAstronomical2024}, with foundation models drawing inspiration from advances in natural language processing and computer vision \cite{rozanskiSpectralFoundationModel2023,leungAstronomicalFoundationModel2024}.
The objective of this work is to explore how deep representation learning can be applied to multimodal stellar spectra to: (i) generate informative representations from varied stellar spectra observations, and (ii) evaluate these representations through three tasks: stellar type classification, regression of stellar parameters, and cross-survey data generation. We also emphasize the scalability of these methods to a range of modalities.
\enlargethispage{2\baselineskip}
To achieve this, we focus on Contrastive Learning (CL) due to its demonstrated effectiveness in downstream tasks across natural language and computer vision, as evidenced by the success of cross-modal models like CLIP \cite{radfordLearningTransferableVisual2021}. CL has also shown promise in galaxy surveys \cite{lanusseAstroCLIPCrossModalPreTraining2023}, is supported by strong theoretical foundations for generating robust representations \cite{wangUnderstandingContrastiveRepresentation2022}, and is well-suited for scalable multimodal learning.

\section{Method: Multimodal Machine Learning and Contrastive Learning}
\label{subsec:Multimodal-Machine-Learning}
\vspace{-1em}
Theoretical considerations attribute multimodal approaches with some advantages relative to unimodal algorithms,
e.g. better training with less data \cite{karchmerStrongerComputationalSeparations2024}.
A common understanding of a \emph{modality} is to associate it with
one specific sensory input, that has been acquired in one specific
way, as opposed to any other. Thus, \emph{multimodality} describes
a research problem or dataset that incorporates multiple different
modalities. Currently, the most prevalent form of multimodality includes
natural language and vision, while historically it's beginnings lie
in audio-visual speech recognition \cite{yuhasIntegrationAcousticVisual1989}. Typical tasks for multimodal ML are: representation,
translation, alignment, fusion or co-learning \cite{baltrusaitisMultimodalMachineLearning2019}, see section \ref{sec:taxonomy} in the appendix for more explanation on these terms. 

\paragraph{Contrastive Learning Implementation}
We implement our model as a coordinated representation by employing one encoder per modality and coordinating the representations via a similarity-based loss (InfoNCE \cite{oordRepresentationLearningContrastive2019}
/ NT-Xent \cite{chenSimpleFrameworkContrastive2020}). Following CLIP \cite{radfordLearningTransferableVisual2021}, we implement CL as in ALIGN \cite{jiaScalingVisualVisionLanguage2021} such that the loss function takes the form:
\begin{align}
\mathcal{L}_{CL} & =-\frac{1}{N}\sum_{i}^{N}log\frac{exp(x_{i}^{T}y_{i}/\tau)}{\sum_{j=1}^{N}exp(x_{i}^{T}y_{j}/\tau)}-\frac{1}{N}\sum_{i}^{N}log\frac{exp(y_{i}^{T}x_{i}/\tau)}{\sum_{j=1}^{N}exp(y_{i}^{T}x_{j}/\tau)}\label{eq:contrastive-loss}
\end{align}
where $x_{i}$ are L2-normalized embeddings of one modality, $y_{i}$
are L2-normalized embeddings of the other modality and $\tau$ is
a temperature hyper-parameter. The second term switches the role of
the modalities in the first term. As eq.~\ref{eq:contrastive-loss} implies, the similarity between samples is calculated as cosine similarity.
The temperature hyper-parameter was introduced in \cite{chenSimpleFrameworkContrastive2020}
to help the model learn from hard negatives - negatives closer to
the anchor than the anchor positive. Working with more than two modalities commonly entails a linear
combination of pairwise contrastive losses.
In detail, for every training batch (16384 spectra), the data is run through the modality specific
encoders and L2-normalized, positioning
the embeddings on a hypersphere. From here, a similarity matrix is
computed, which calculates the dot-product of every cross-modality
instance combination scaled by the temperature $\tau$ from eq.~\ref{eq:contrastive-loss}. 
The diagonal elements correspond to positive
pairings; every other element to a negative pairing.
The loss (eq.~\ref{eq:contrastive-loss}) can be calculated by applying
multi-class softmax cross-entropy (also called categorical cross-entropy) to every row, where only the diagonal entries (positive pairs) are
considered as correct class predictions. 
To calculate the symmetrizing second term of eq.~\ref{eq:contrastive-loss},
the same calculation is repeated with a transposed similarity matrix and both terms are added for the total loss.
We use the LAMB optimizer \cite{youLargeBatchOptimization2020} because of its higher performance capabilities on bigger
batches. At the start of every epoch, the training set is shuffled to avoid that any data instance would
always only be compared to a fixed subset (the corresponding batch)
of other instances. We choose a fixed temperature of $\tau=0.01$ as in \cite{lanusseAstroCLIPCrossModalPreTraining2023} (but see also \cite{radfordLearningTransferableVisual2021} for a trainable temperature). Note, the results presented here depend on the exact temperature value chosen. We have explored different temperature values and found $\tau=0.01$ to perform best. For the latent size, the physical dimensionality is taken into account--eight stellar parameters and up to 20 abundances--which we round
to the next highest power of two, i.e. 32. 

\paragraph{Network Architectures}
Two different data formats are encountered -- a spectrum and spectral coefficients. 
The physics of spectra suggests strong correlations between adjacent bins. RVS spectra are encoded by convolutional neural networks (CNNs) while spectral coefficients
are encoded by a 1-layer MLP. 
\textit{The 1-layer MLP} consists of an input layer, one hidden layer and an output layer, the biases of the linear combinations are set to zero and the
non-linear function is Leaky-ReLu.
Additionally, to combat overfitting, Dropout is used between the input
and hidden layer with dropout probability $p_{\mathrm{dpo}}=0.2$. Dropout is only
active in the training phase, in the evaluation phase no output is
masked. Entries, which are not masked out are scaled by $1/(1-p_{\mathrm{dpo}})$
to keep the expected output for training the same as for testing.
After an extensive hyper-parameter search we choose
for the RVS~1-Layer~MLP a hidden-layer size of 8192 and for the XP~1-Layer~MLP 1024. 
Our \textit{Convolutional Neural Network}
is adopted from the RVS-CNN by \cite{guiglionGaiaDR3Tracing2024}.
Again, the linear combination is
implemented without bias terms and the beginning and end
of the layer inputs are padded by zeros, to allow the kernel to center
on the positions of the edge elements. Convolutions are implemented in parallel, which in turn increases the channel
dimensionality. 
Dimensionality reduction is achieved by
max-pooling layers. 
Every layer, except the output and max pooling layers, implements
Leaky-ReLu as activation. The final feed-forward network implements
layer-wise dropout with a dropout rate of 20\%. See \cite{guiglionGaiaDR3Tracing2024} for details on the network.
\enlargethispage{\baselineskip}
Our \textit{cross-modal generation decoder} follows 
\cite{melchiorAutoencodingGalaxySpectra2023}
which implement a MLP decoder for galaxy spectra which we transfer to the RVS spectra.
Key feature is an activation function of the
form
\begin{equation}
f(x)=\left[\gamma+\frac{1-\gamma}{\left(1+\mathrm{e}^{-\beta\odot x}\right)}\right]\odot x
\end{equation}
where $\gamma$ and $\beta$ are trainable parameters \citep{alsingSPECULATOREmulatingStellar2020}. This activation is able to cover smooth features
for small $\beta$ and sharp changes in gradient for $\beta\to\infty$ which supports easier modeling of spectral
lines. For XP coefficients we use a simple MLP.

\paragraph{Code and Hardware}
All code is implemented in the python library \texttt{JAX} \cite{jax2018github}. 
Neural networks are build with the \texttt{Flax} \cite{flax2020github} package. Training routines, like optimizer and learning rate scheduling, are supported by \texttt{Optax}, while saving and loading network parameters is done by \texttt{Orbax} \cite{deepmind2020jax}. Datasets are handled by the Huggingface \texttt{datasets} package \cite{lhoest-etal-2021-datasets}. All our code will be publicly available upon acceptance.
Training and inference takes place on a single Nvidia A100 with 40~GB of VRAM.

\paragraph{Dataset}
We use Gaia DR3 RVS spectra and Gaia XP coefficients of 841,300 instances from \cite{guiglionGaiaDR3Tracing2024}. This dataset is ideally suited for the task of exploring representation learning as the RVS spectra represent a relatively small dataset of high quality and high information content while the XP coefficients encompass a much larger dataset of low quality data. Being able to transfer labels and find a common representation space for these two data modalities nicely showcases the benefit of representation learning for stellar spectra analysis.

For an additional 44,780 samples, APOGEE labels are available. 
In addition, spectral types are added from Gaia's Extended Stellar Parametrizer for Emission-Line Stars (ESP-ELS) \cite{fouesneauGaiaDataRelease2023}. The RVS spectra have been z-scale normalized to facilitate training the neural networks. Additionally, all BP \& RP coefficients have been divided by the first coefficient, like in \cite{guiglionGaiaDR3Tracing2024}.
Afterwards, the first coefficients of BP \& RP have been log-scaled to bring the distribution closer to a normal distribution, and then z-scaled as well. All other XP entries are left in their unnormalized state, as their absolute magnitudes contain information on their relevance. 
The 841,300 instances without APOGEE labels are used as the training set. While, the 44,780 instances with APOGEE labels constitute the validation set, to facilitate downstream regression tasks for validation. 

\section{Results}

\paragraph{Structured Embedding Space\protect\label{sec:test-viz-emb}}
\begin{wrapfigure}[20]{R}{0.5\textwidth}
\centering
\vspace{-1.75cm}
\includegraphics[width=0.495\textwidth]{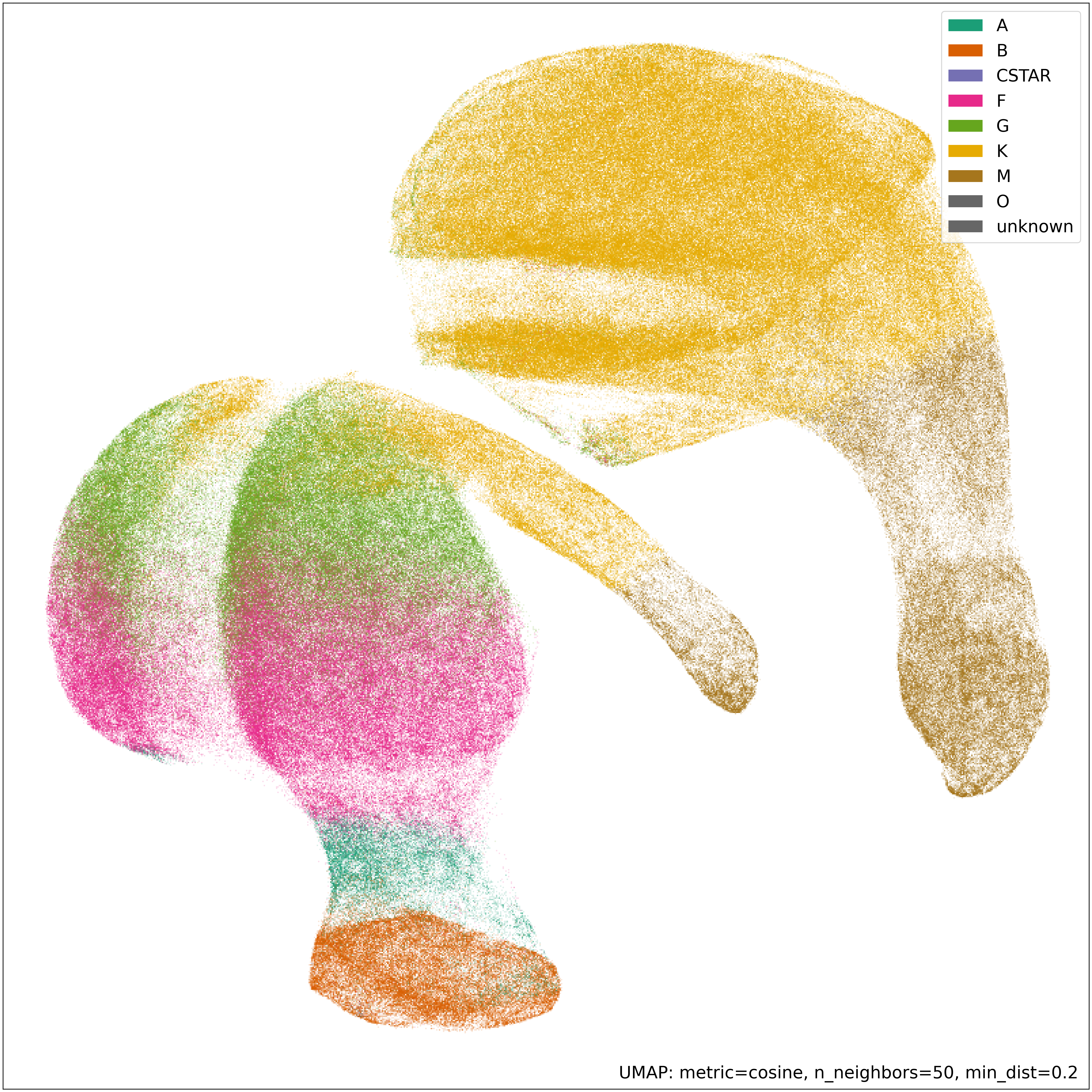}
\vspace{-.25cm}
\caption{UMAP visualization of the RVS training set embeddings with parameters \texttt{metric=cosine}, \texttt{n\_neighbors=50}, \texttt{min\_dist=0.2}; colored by spectral type.}
\label{fig:test-viz-class}
\end{wrapfigure}
For a qualitative inspection of the information content and structuredness of the embedding space we apply UMAP \cite{mcinnesUMAPUniformManifold2020}
 to generate a 2-d visualization of the 32-d embedding space. 
We color the resulting UMAP embedding for the RVS training data by spectral type (Fig.~\ref{fig:test-viz-class})
and stellar parameters $T_{\mathrm{eff}}$, $\log g$ and the abundance of all metals ($[M/H]$) and $\alpha$-elements ($[\alpha/M]$, Fig.~\ref{fig:test-viz-class2} in the appendix). We have further analysed the XP validation set (see Fig.~\ref{fig:test-viz-class2} in the appendix) and find the same results as for RVS training data which we show here.
All projections colored by spectral type show large
homogeneously colored regions. This implies that in the original embedding
space, objects with the same type share a common neighborhood. Moreover,
some classes continuously transition into others, like K \& M, F \&
G, B \& A. These pairings also form more separated clusters from another.
Thus leading to separations from G to K and A to F. Additionally,
there is some kind of duplication, in that there are two big clusters
with a similar color sequence and subclusters. Overall, it is noteworthy
that the transitions between classes coincide with the physically
informed stellar sequence: OBAFGKM.
An inspection of the projections colored by stellar parameters (see Fig.~\ref{fig:test-viz-rvs-train-params} in the appendix) shows
continuous transitions between parameter values. This is most notable
for the effective temperature $T_{\mathrm{eff}}$, the surface gravity
$\log g$ and the abundance of $\alpha$-elements $[\alpha/M]$. Still,
also for $[M/H]$ homogeneous color patches are visible, if slightly
more interspersed in some regions for the training set.  Additionally, we find two separate patches where $T_{\mathrm{eff}}\sim4500$ K (spectral class M) is similar but $\log g$ is different (2 vs. 4) which might point to a physical separation of dwarfs and giants. All together,
the UMAP projections indicate, that the embedding space encodes physically meaningful information.

\paragraph{Zero-shot regression with k-Nearest Neighbours\protect\label{subsec:test-kNN}}

\begin{figure}
\centering
\vspace{-.85cm}
\includegraphics[width=.49\textwidth]{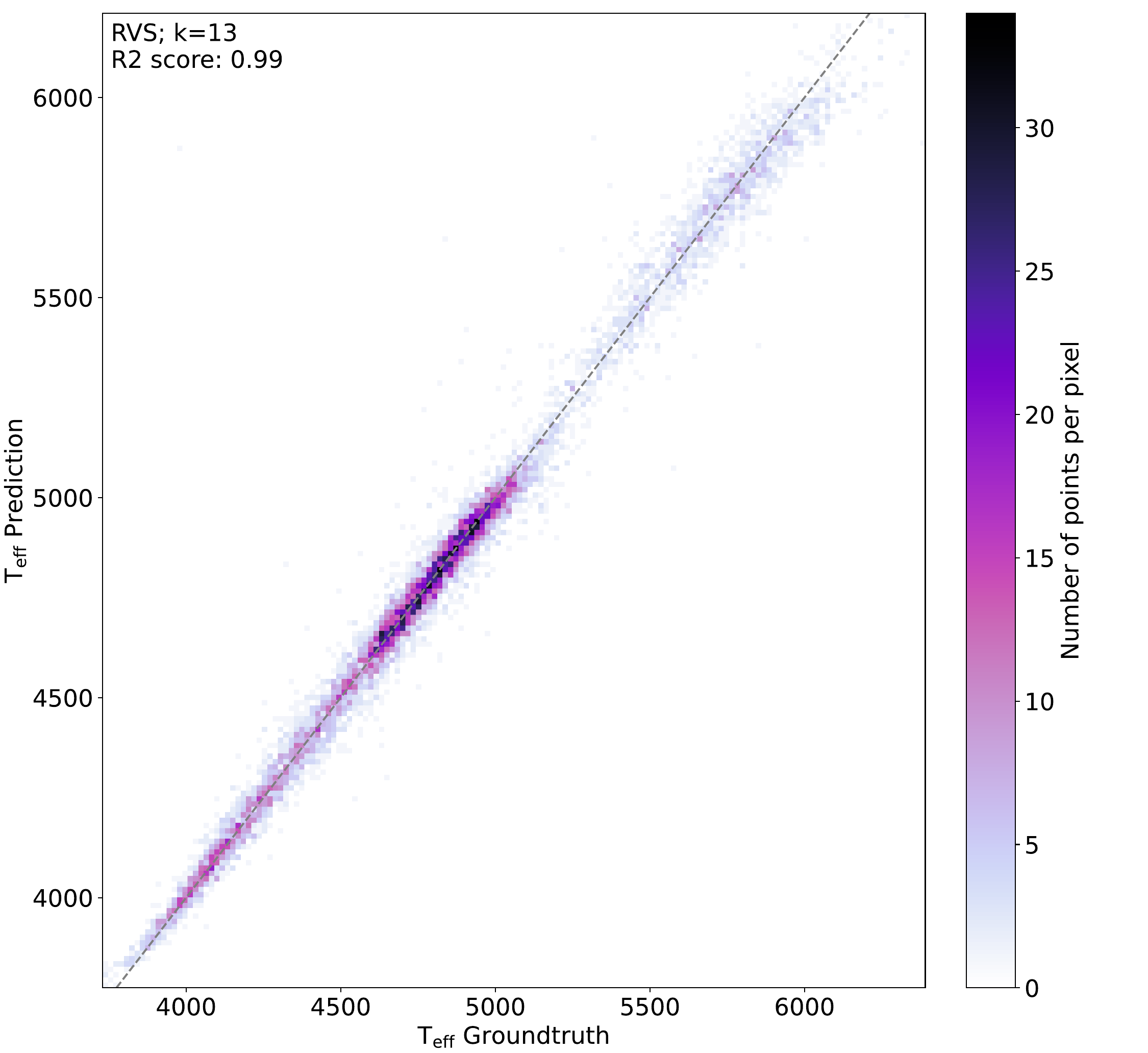}
\includegraphics[width=0.49\textwidth]{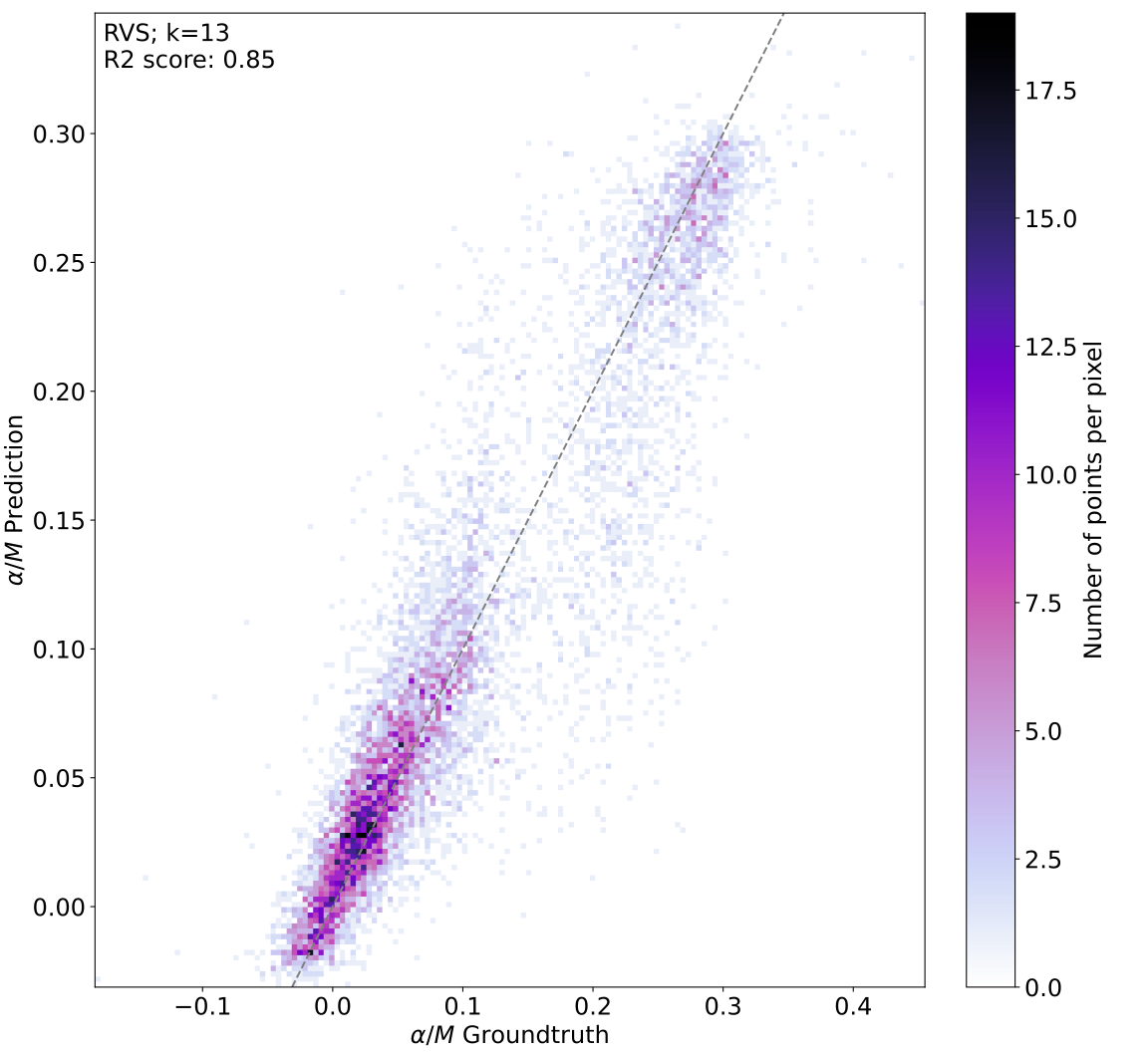}
\caption{Prediction-vs-ground-truth plots for k-NN predictions (k=13) from RVS and XP for $T_{\mathrm{eff}}$ and $[\alpha/M]$; grey line corresponds to the one-to-one line.}
\label{fig:test-knn-reg-vs}
\end{figure}

Next, zero-shot regression onto stellar parameters is performed and the $\mathrm{R}^{2}$-score, also called coefficient of
determination, for each of the stellar parameters is calculated. In Fig.~\ref{fig:test-knn-reg-vs} we show results for $T_{\mathrm{eff}}$ and $[\alpha/M]$ for $k=13$. The effective temperature achieves
the highest score with 0.9874, while the $\alpha$-element
abundance performs worst with a $\mathrm{R}^{2}$-score of 0.8488. 
We find that $T_{\mathrm{eff}}$ shows only a small spread over
the whole temperature domain, while $[\alpha/M]$ exhibits larger
errors between 0.1 and 0.2~dex. In general a high $\mathrm{R}^{2}$-score from a k-NN algorithm was again to be expected, since the visual inspection in \ref{fig:test-viz-rvs-train-params} also indicated homogeneous neighbourhoods with respect to stellar parameters.

\paragraph{Cross-modal translation}
The four closest entries based on the cosine similarity are
retrieved to perform cross-modal translation. In the appendix, Fig.~\ref{fig:test-knn-lookup-rvs}
shows RVS spectra retrieved by XP coefficients, and Fig.~\ref{fig:test-knn-lookup-xp}
XP coefficients retrieved by RVS spectra. Note, that in the case of
spectra the retrieved neighbors show variation in the noisy plateau
area. Meanwhile, in the area of spectral lines the retrieved entries
are mostly placed directly behind the ground truth. For the XP coefficients
the variation is higher, with some neighbors in proximity of the ground truth,
but not all.

\begin{figure}
\centering{}\includegraphics[width=\textwidth,trim={0cm 0cm 0cm 0cm},clip]{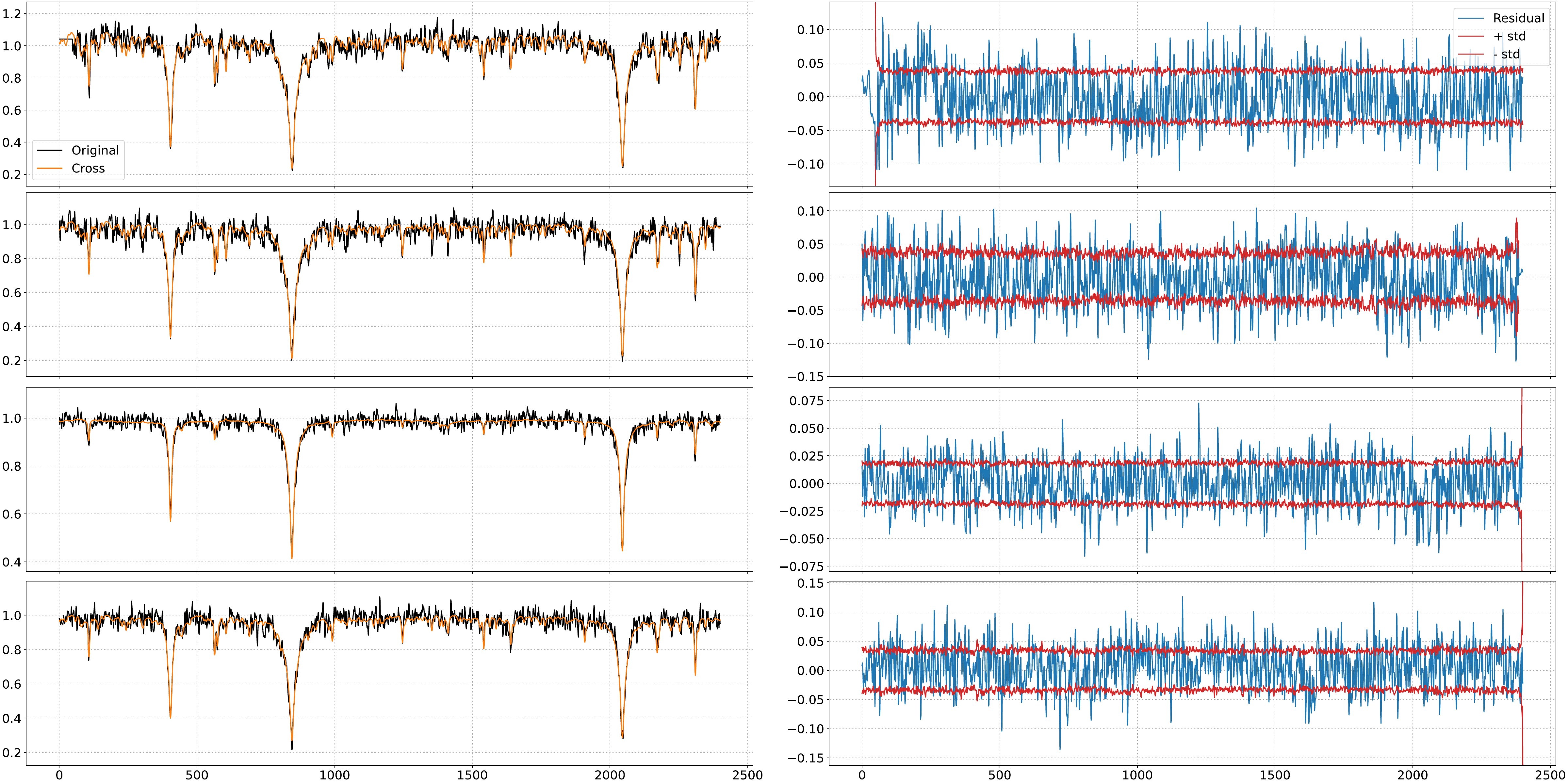}
\caption{Four samples of cross-modal generation with fusion scheme Single Fusion; on the right side the residuals are plotted against the ground truth measurement error.}
\label{fig:test-fusion-gen-samples}
\end{figure}


\paragraph{Cross-modal generation}
Using the cross-modal decoder, we investigate the performance of cross-modal generation of the more complicated modality (RVS) from the simpler one (XP).
One example RVS spectra generated from XP coefficient embeddings together with the ground truth spectra and the respective residual is shown in Fig.~\ref{fig:test-fusion-gen-samples} and more examples can be found in the appendix in Fig.~\ref{fig:fusion-gen-samples}. This analysis shows that the residuals are on the magnitude of the measurement error and that key features of the spectra such as strong and weak absorption lines are well reproduced.

\section{Summary and Outlook\protect\label{chap:Summary-and-Outlook}}
\vspace{-1em}
In this work, we have used contrastive learning, a state-of-the-art multimodal ML algorithm, to generate informative representations of stellar spectra from multi-modal data, namely Gaia RVS and XP spectra.
We map the raw data into a shared representation space using a CNN for RVS and a 1-layer MLP for XP.
We find that multi-modal learning creates a highly structured latent embedding of the stellar spectra that aligns well with fundamental stellar parameters. We show the information content of this embedding space for downstream task such as regression, classification and cross-modal
translation via simple k-Nearest Neighbor search.
Moreover, we explore cross-modal generation via decoder networks to generate RVS spectra from XP coefficients and vice-versa.
For all tasks we find excellent performance which highlights the benefit of contrastive learning for stellar spectra analysis from multiple different surveys.


In general, taking on more modalities and leveraging on the large body of stellar spectral surveys is the obvious next direction of this project.
Another avenue, would be to train not only observed
spectra, but also on synthetic ones. This could also be combined with pre-training an encoder on synthetic spectra, to then only fine-tune the last layers for the real spectra.
Regarding the encoder networks themselves, several other networks might be explored.
Specifically, the XP coefficients should be better processed by an attention-based
architecture. This is speculated because the spectral coefficients
don't share stronger correlations with their neighbors. 
However, spectral data can also be processed by attention-layers, after CNNs have reduced the dimensionality of the input.
Lastly, the cross-modal generation of any data is restricted by the
capabilities of the decoder. Recently, the design which showed the
best performance for the generation task are diffusion models. For
those, Contrastive Learning lays the foundation, in that the learned
embeddings are used as conditions for generation. Furthermore, diffusion
models often employ a pre-trained autoencoder. Since in the context
of this work contrastive learning, as well as autoencoder training
routines were implemented, the way towards diffusion models in multimodal
astrophysics has been paved.
%
Finally, to extend this work to probabilistic representation spaces, entails expanding the similarity metric to compare probabilistic embeddings.


\section*{Broader impact statement}
The authors are not aware of any immediate ethical or societal implications of this work. This work purely aims to aid scientific research and proposes to apply multi-modal contrastive learning techniques to stellar spectra to learn about fundamental physics.

\begin{ack}
This work is funded by the Carl-Zeiss-Stiftung through the NEXUS programm.
\end{ack}


\bibliography{bib.bib}

\appendix

\section{Appendix / supplemental material}

\subsection{Multi-modal learning taxonomy}
\label{sec:taxonomy}

\begin{description}
\item [{Representation}] Any vector or tensor representing an entity, is
referred to as a representation (equivalently embedding)
\cite{bengioRepresentationLearningReview2014}. For a multimodal problem
the challenge lies in exploiting the complementarity and redundancy
of multiple modalities which is complicated by the heterogeneity
of the multimodal data.
\item [{Translation}] Given an entity in one modality the task is to generate
the same entity in a different modality. The challenge arises from
the possible existence of multiple mapping outputs that are still
consistent/correct.
\item [{Alignment}] The task is to find direct relations and correspondences between sub-components of several modalities, by some kind of similarity
definition. 
\item [{Fusion}] This task entails integrating data from multiple modalities
with the goal of predicting an outcome measure, e.g. performing classification
or regression. The challenge is posed by each modality's varying predictive
power or noise regarding the outcome. Also, at least one modality
might be missing at inference time. Fusion benefits from multimodality,
since predictions become more robust, complementary information are
utilized and the model gains redundancy regarding sensory absence.
\item [{Co-Learning}] Here, the modeling of a (resource poor) modality
is aided by exploiting knowledge from another (resource rich) modality.
Limited resources might entail: lack of annotated data, noisy input,
and unreliable labels. ``Learning'' points to the helper modality
only being used during training, not testing time.
\end{description}

\section{Additional Figures}

\begin{figure}
\centering{}%
\includegraphics[width=0.48\textwidth]{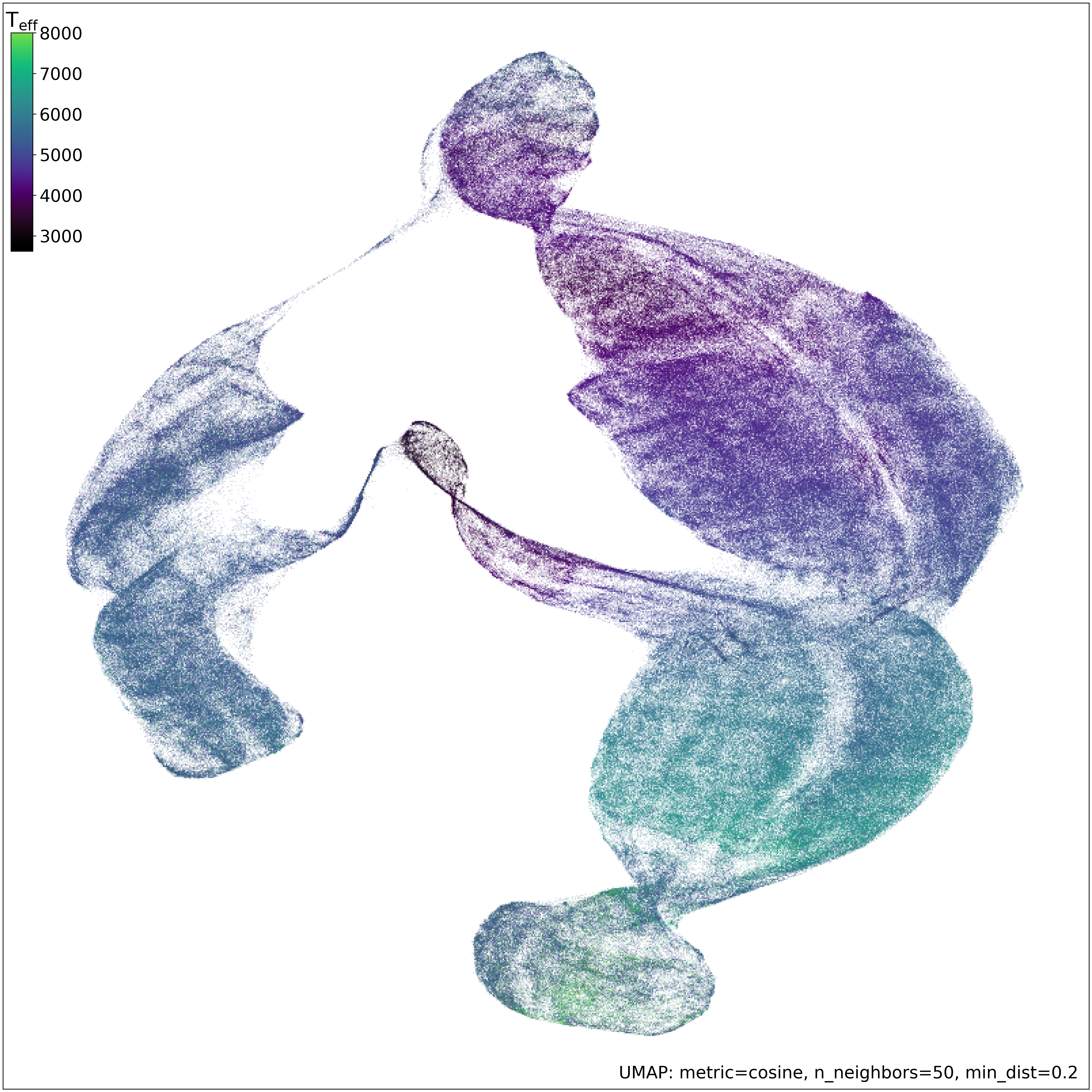}
\includegraphics[width=0.48\textwidth]{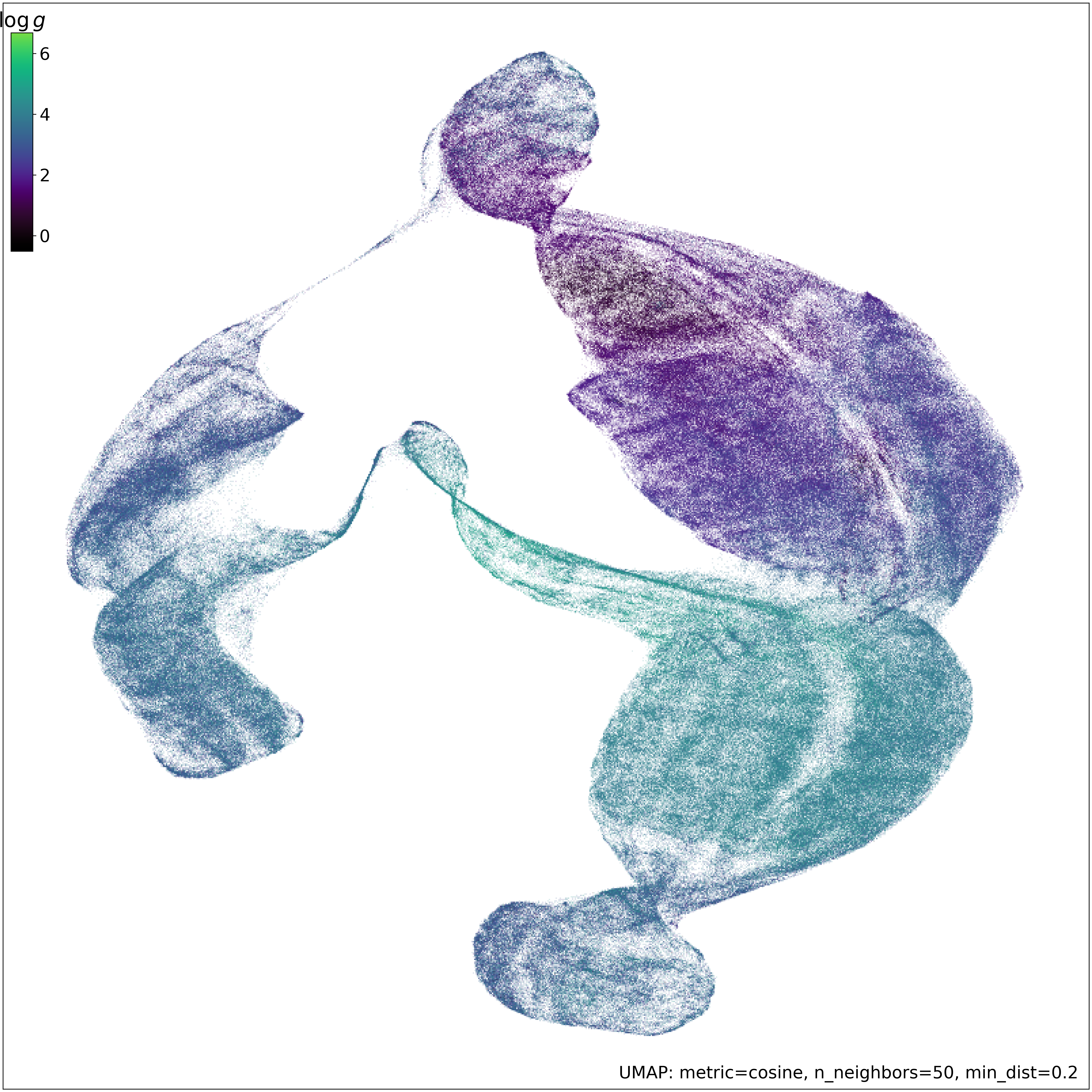}
\includegraphics[width=0.48\textwidth]{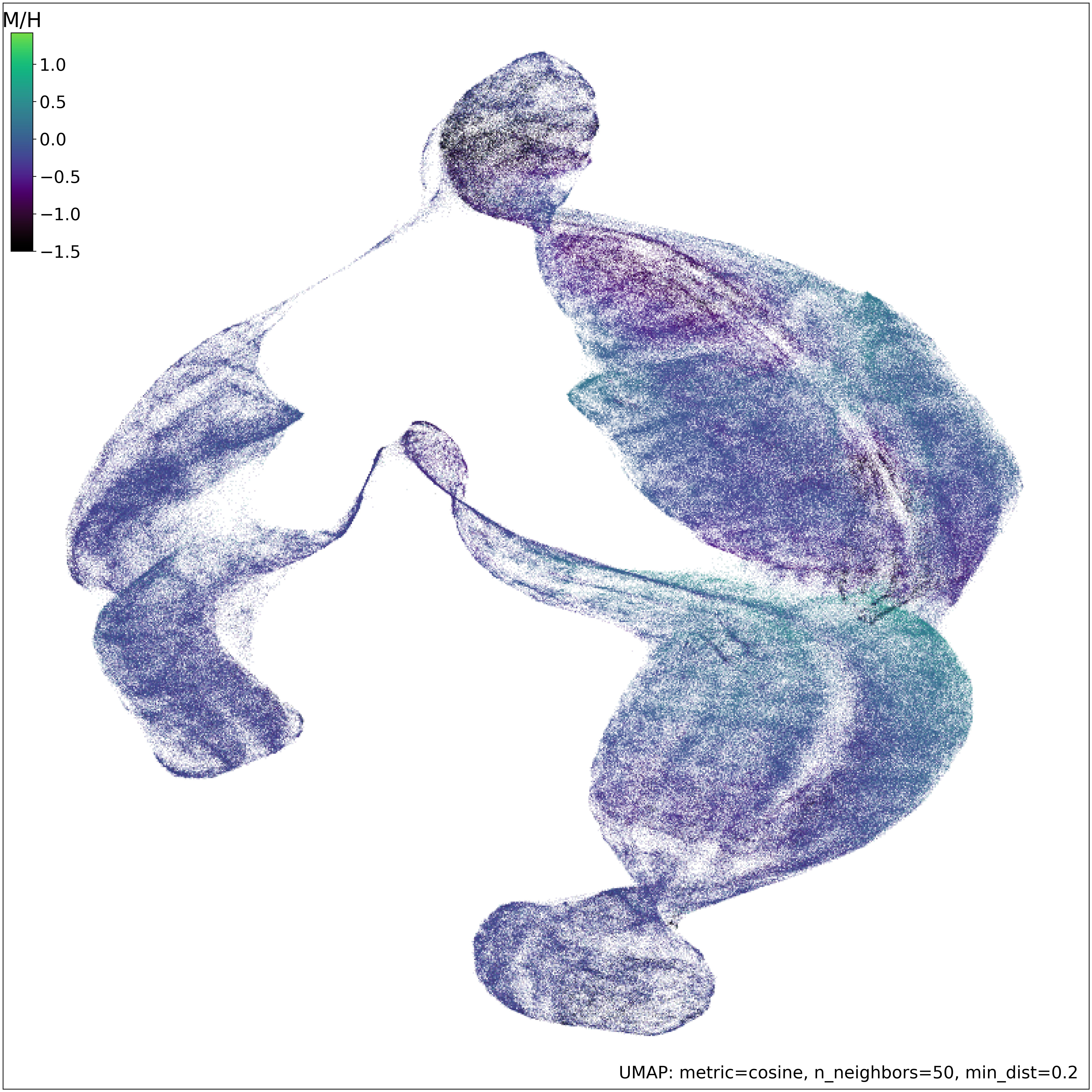}
\includegraphics[width=0.48\textwidth]{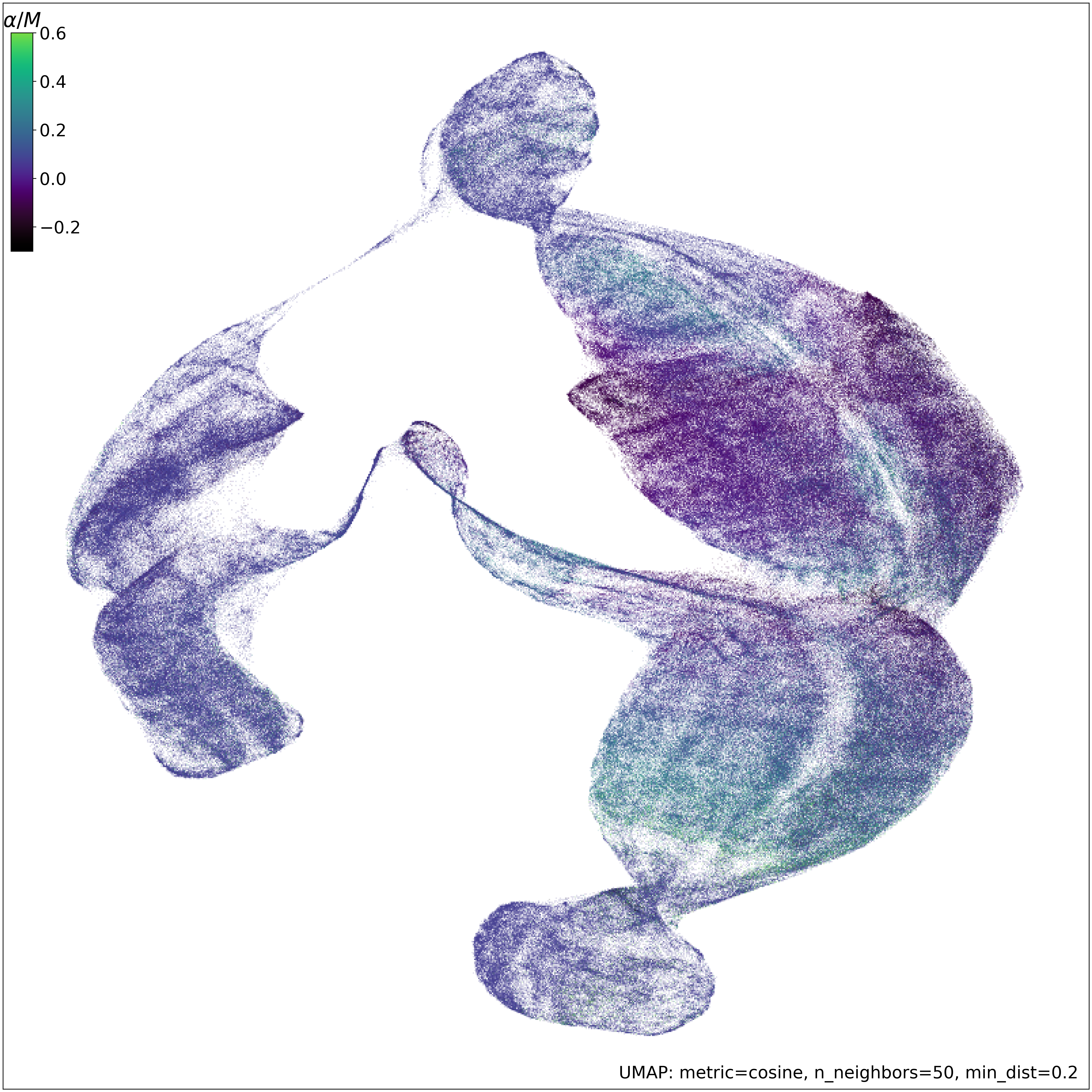}
\caption{UMAP visualization of RVS training set embeddings with parameters
\texttt{metric=cosine}, \texttt{n\_neighbors=50}, \texttt{min\_dist=0.2}; colored by stellar parameters; $T_{\mathrm{eff}}$ upper left, $\log g$ upper right, the abundance of all metals ($[M/H]$) lower left and $\alpha$-elements $[\alpha/M]$ lower right.}
\label{fig:test-viz-rvs-train-params}
\end{figure}



\begin{figure}
\centering
\includegraphics[width=0.48\textwidth]{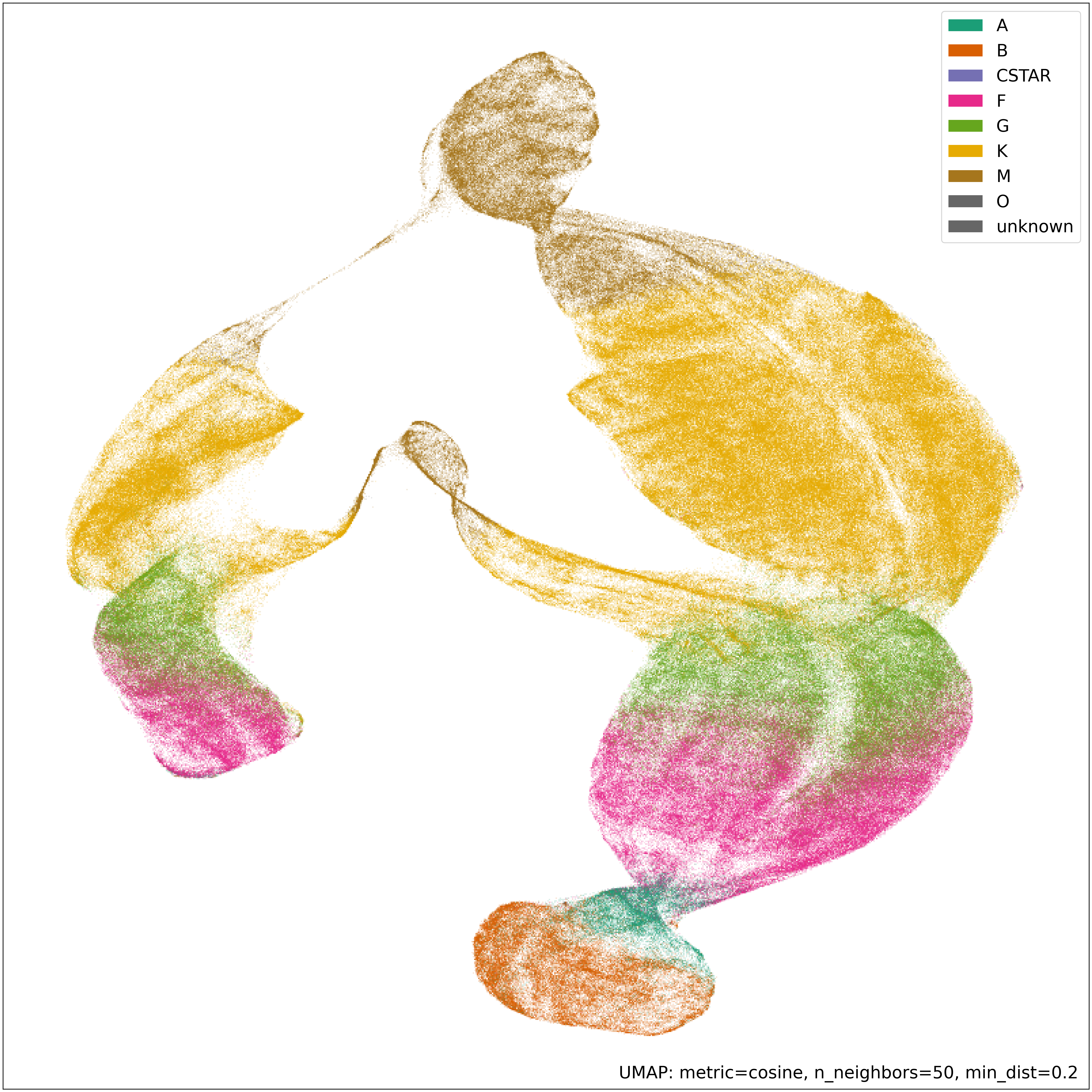}
\caption{UMAP visualization of the XP validation set embeddings with parameters \texttt{metric=cosine}, \texttt{n\_neighbors=50}, \texttt{min\_dist=0.2}; colored by spectral type.}
\label{fig:test-viz-class2}
\end{figure}


\begin{figure}
\begin{centering}
\includegraphics[width=0.95\textwidth]{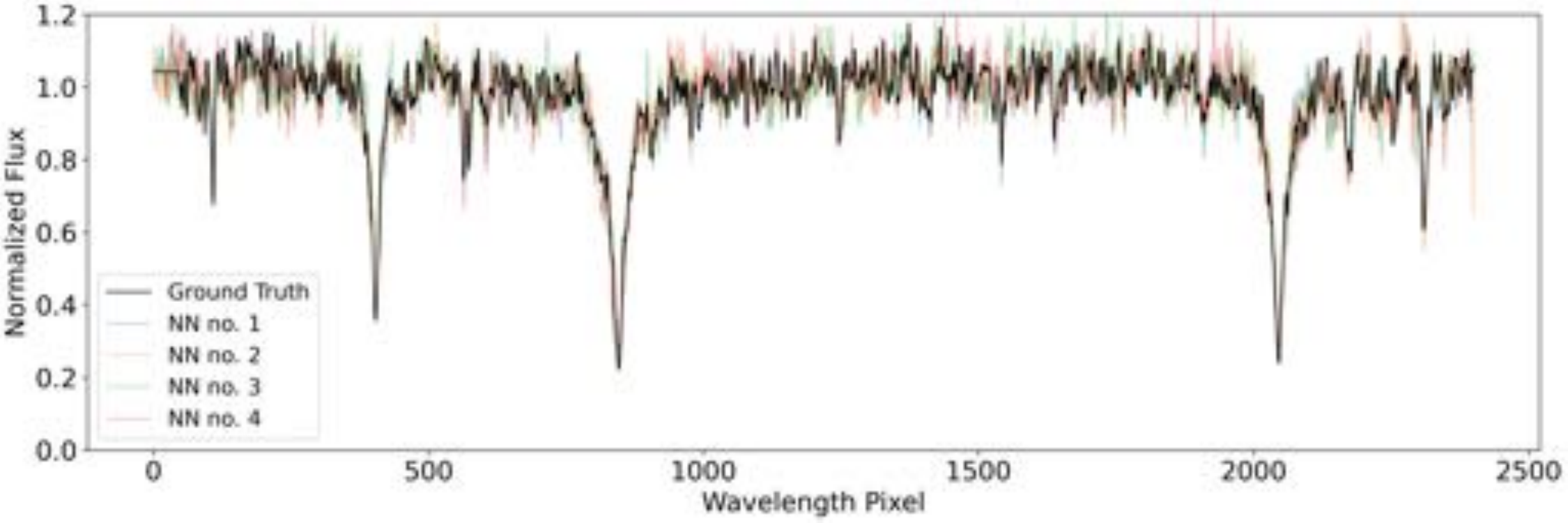}\vfill{}
\includegraphics[width=0.95\textwidth]{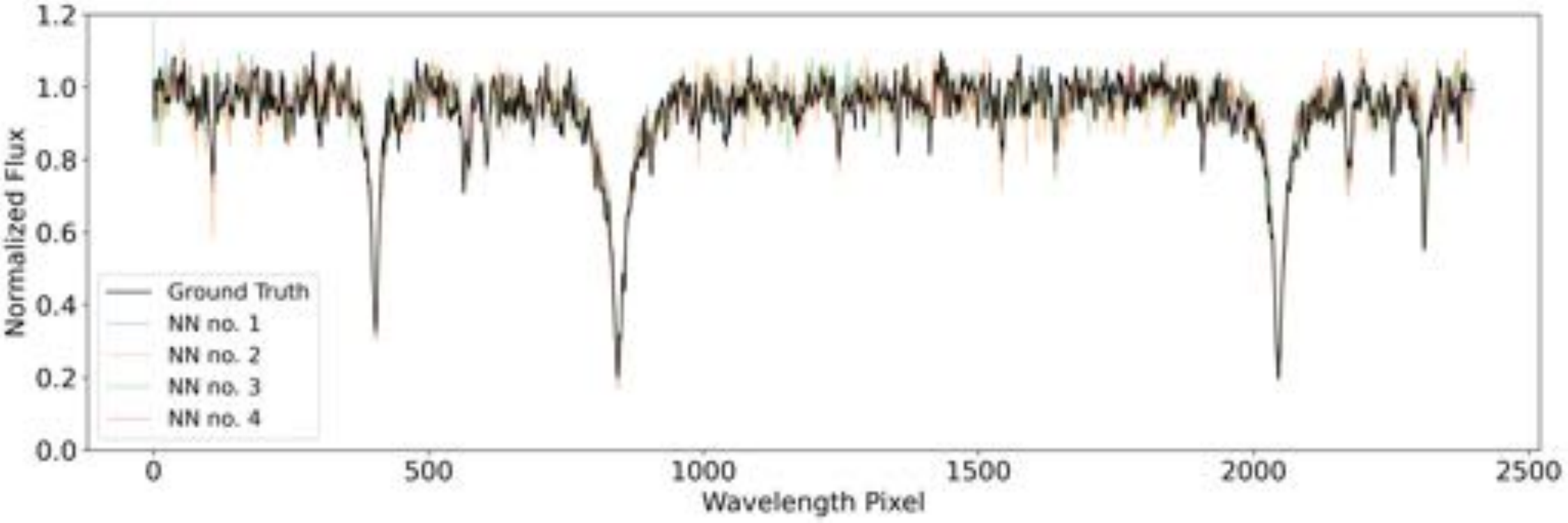}\vfill{}
\includegraphics[width=0.95\textwidth]{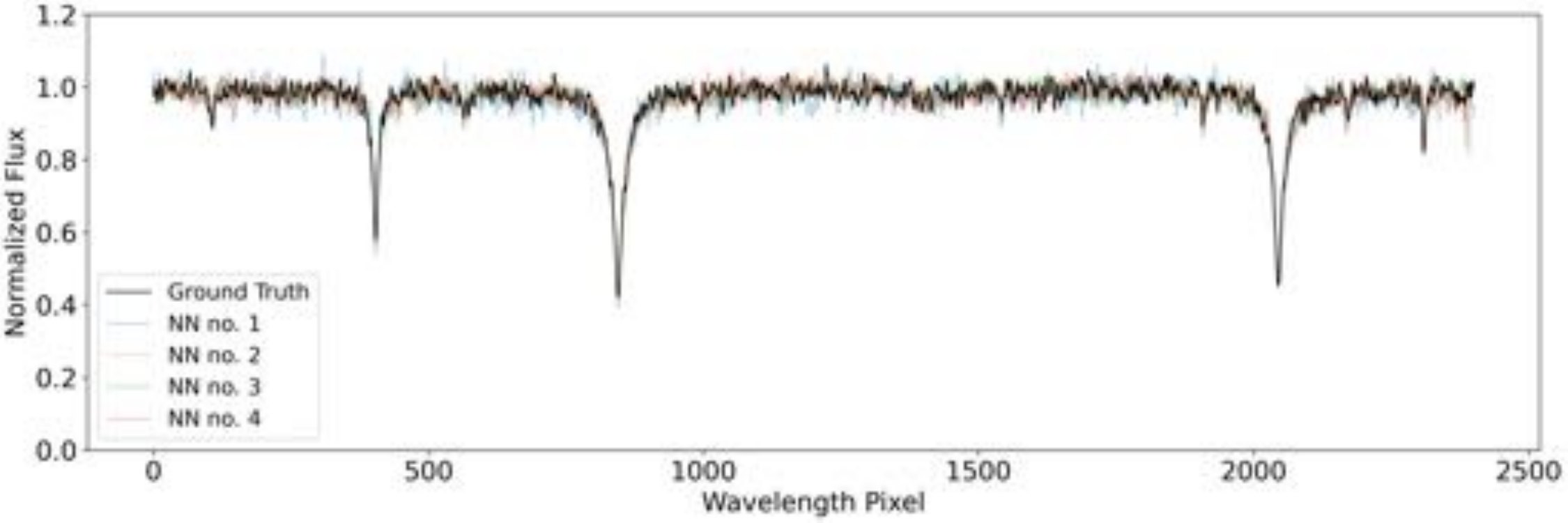}\vfill{}
\includegraphics[width=0.95\textwidth]{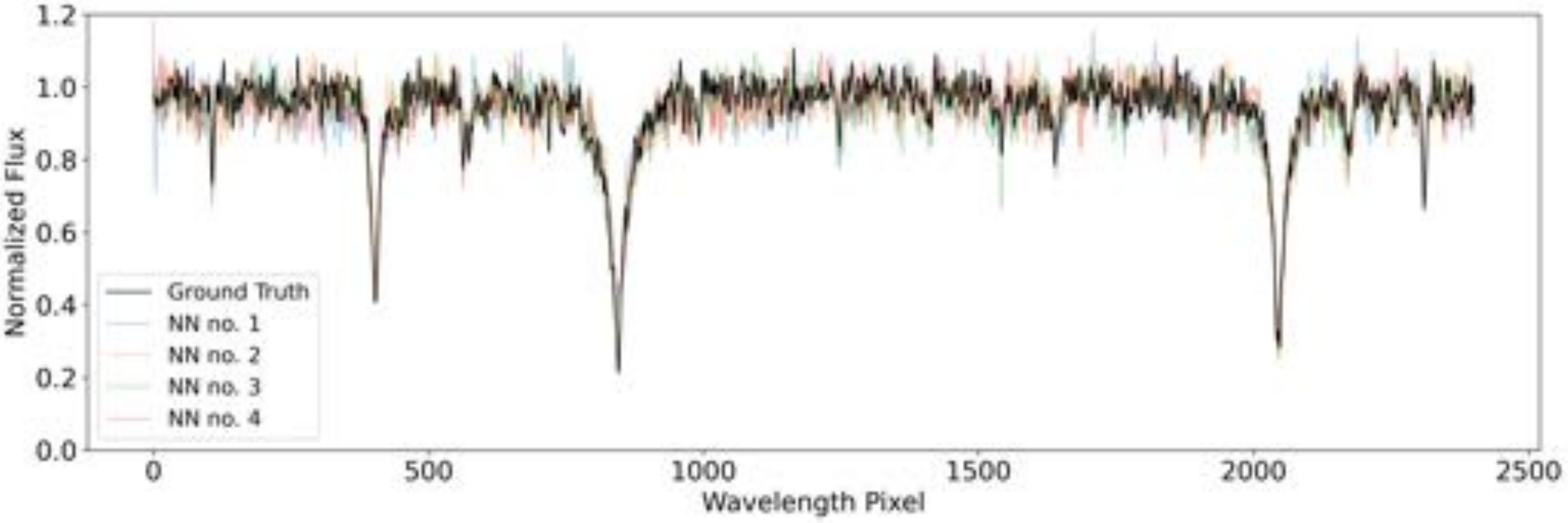}
\par\end{centering}
\caption{Similarity lookup from XP coefficients to RVS spectra for four example coefficients.}
\label{fig:test-knn-lookup-rvs}
\end{figure}

\begin{figure}
\begin{centering}
\includegraphics[width=0.9\textwidth]{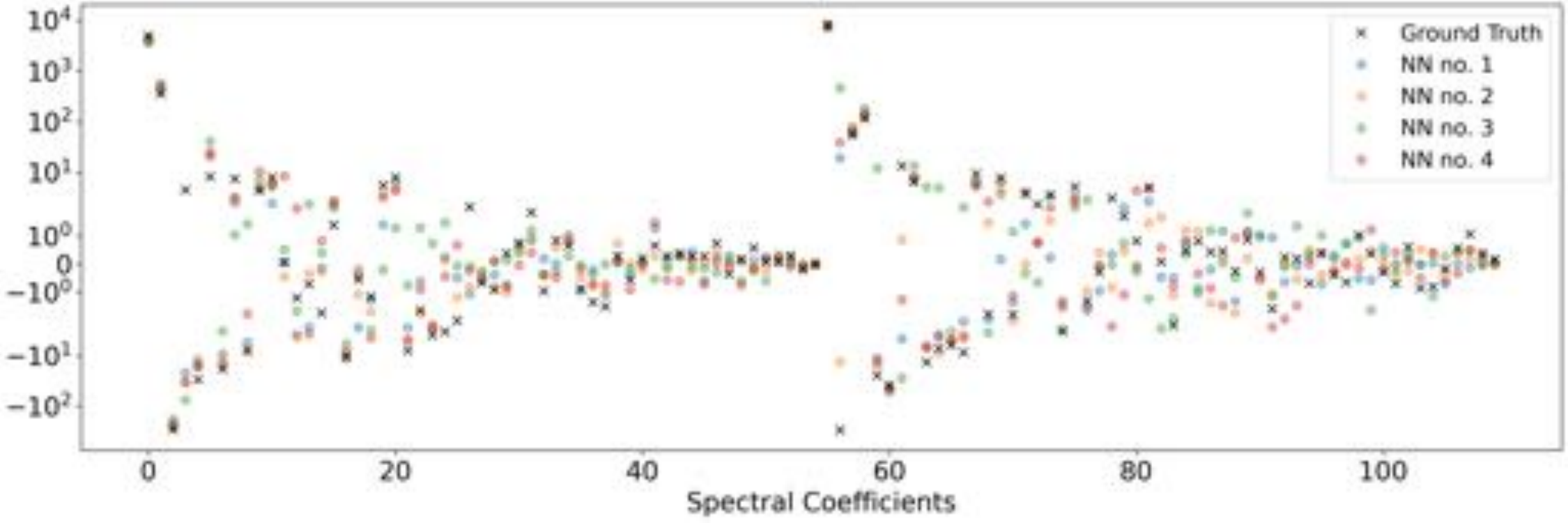}\vfill{}
\includegraphics[width=0.9\textwidth]{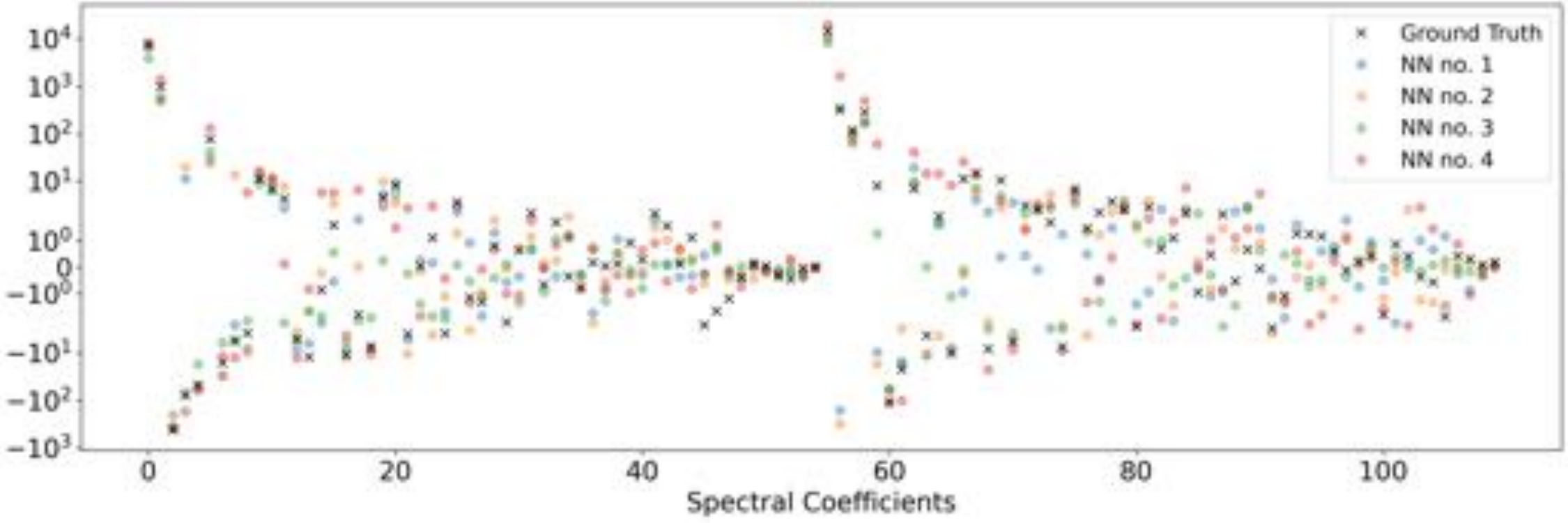}\vfill{}
\includegraphics[width=0.9\textwidth]{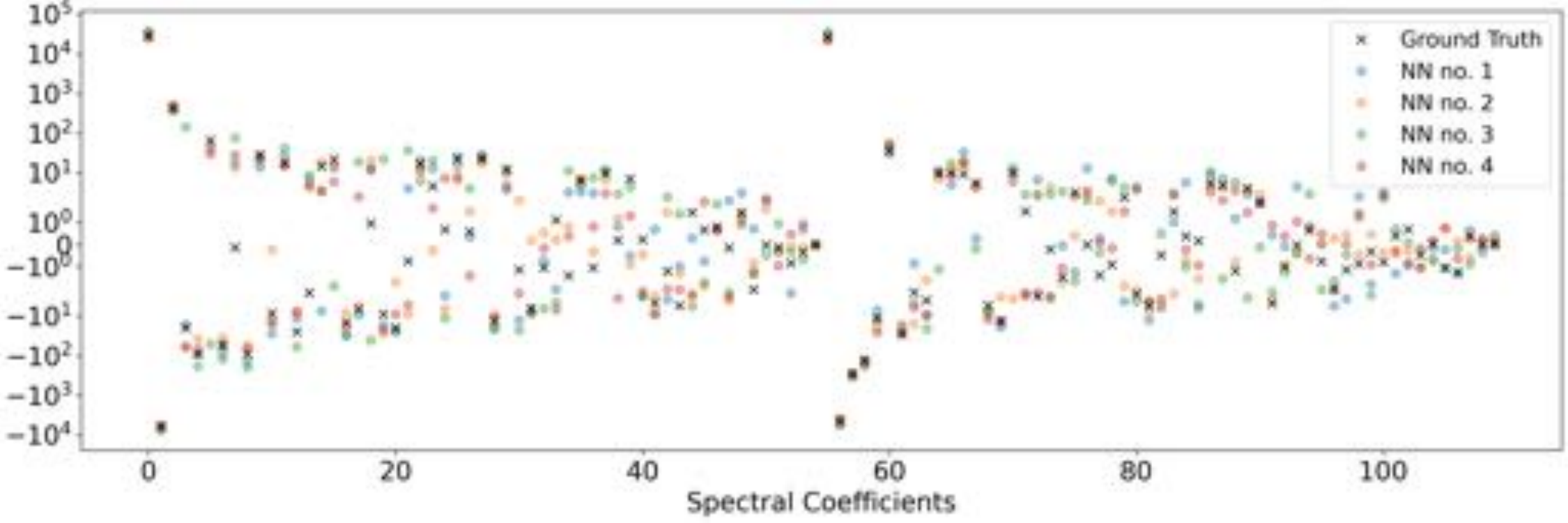}\vfill{}
\includegraphics[width=0.9\textwidth]{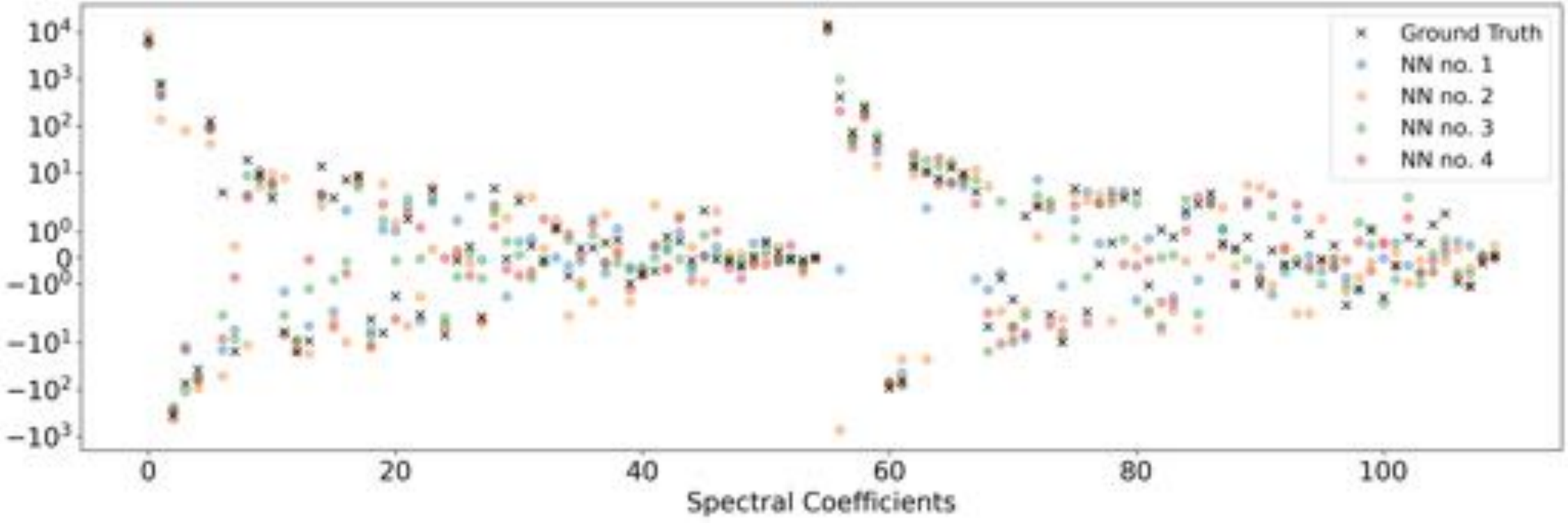}
\par\end{centering}
\caption{Similarity lookup from RVS spectra to XP coefficients for four example spectra; the scale is
a symmetric logarithmic scale with linear scaling around zero.}
\label{fig:test-knn-lookup-xp}
\end{figure}


\end{document}